\documentclass[11pt]{article}
\pdfoutput=1
\usepackage{jcapmod}
\usepackage{booktabs}
\usepackage[english]{babel}
\usepackage{amsmath,amssymb,amsbsy,amstext, amsthm, simplewick}
\usepackage{graphicx}
\usepackage{amsfonts}
\usepackage{amssymb}
\usepackage{bbm}
\usepackage{upgreek}
 \usepackage{exscale,relsize}
 \usepackage[makeroom]{cancel}
\usepackage{soul}
\usepackage{bm}

\RequirePackage{color}

\usepackage{colortbl}
\definecolor{rp}{cmyk}{0.2, 1, 0.6, 0}
\definecolor{green2}{cmyk}{0, 1, 0.5, 0}
\definecolor{lightgreen}{cmyk}{0.2, 0, 0.2, 0.2}
\definecolor{lightgray}{cmyk}{0.1,0.2,0,0.1}
\definecolor{lightgray2}{cmyk}{0.4,0.4,0,0.8}
\definecolor{black}{cmyk}{1.0,1.0,1.0,1.0}

\allowdisplaybreaks[1]

\usepackage{colortbl}
\definecolor{lightgreen}{cmyk}{0.2, 0, 0.2, 0.2}
\definecolor{lightgray}{cmyk}{0.1,0.2,0,0.1}
\definecolor{lightgray2}{cmyk}{0.1,0.1,0,0.1}

\makeatletter
\newlength{\apb@width}
\newcommand{\autoparbox}[2][c]{\settowidth{\apb@width}{#2}\parbox[#1]{\apb@width}{#2}}

\makeatother

\numberwithin{equation}{section}

\def\beq{\begin{equation}}
\def\eeq{\end{equation}}

\def\bea{\begin{eqnarray}}
\def\eea{\end{eqnarray}}

\def\dd{{\rm d}}
\def\Tr{{\rm Tr}}

\def\beq{\begin{equation}}
\def\eeq{\end{equation}}
\def\bea{\begin{eqnarray}}
\def\eea{\end{eqnarray}}

\def\dd{{\rm d}}
\def\Tr{{\rm Tr}}

\def\0{{\boldsymbol 0}}

\def\Tr{{\textrm{Tr}}}

\DeclareRobustCommand{\SkipTocEntry}[4]{}

\begin{document}

\begin{titlepage}

\setcounter{page}{1} \baselineskip=15.5pt \thispagestyle{empty}

\bigskip\

\vspace{1cm}
\begin{center}

{\fontsize{20}{24}\selectfont  \sffamily \bfseries  Vacuum Decay in CFT}\vskip 5 pt
{\fontsize{20}{24}\selectfont  \sffamily \bfseries  and the Riemann-Hilbert problem}\vskip 5 pt

\end{center}

\vspace{0.2cm}
\begin{center}
{\fontsize{13}{30}\selectfont Guilherme L. Pimentel$^{\bigstar}$, Alexander M. Polyakov$^{\blacklozenge}$ and Grigory M. Tarnopolsky$^{\blacklozenge}$} 
\end{center}

\begin{center}

\vskip 8pt
\textsl{$^\bigstar$ Department of Applied Mathematics and Theoretical Physics, Cambridge University} \vskip 1pt \textsl{Cambridge, CB3 0WA, UK}
\vskip 7pt

\textsl{$^\blacklozenge$ Joseph Henry Laboratories, Princeton University}
\vskip 1pt \textsl{Princeton, NJ 08544, USA}

\end{center}

\vspace{1.2cm}
\hrule \vspace{0.3cm}
\noindent {\sffamily \bfseries Abstract} \\[0.1cm]
We study vacuum stability in $1+1$ dimensional Conformal Field Theories with external background fields. 
We show that the vacuum decay rate is given by a non-local two-form. This two-form is a boundary term that must be added to the effective in/out Lagrangian. The two-form is expressed in terms of a Riemann-Hilbert decomposition for background gauge fields, and its novel ``functional'' version in the gravitational case.

\vskip 10pt
\hrule
\vskip 10pt

\vspace{0.6cm}
 \end{titlepage}

\newpage
\section{Introduction}

In this article we discuss vacuum decay in $1+1$ dimensional Conformal Field Theories with external fixed background fields. As an example, we consider a theory of massless fermions in $1+1$ dimensions coupled to Abelian, non-Abelian or gravitational background fields.  The computation of the vacuum decay rate involves evaluating the effective action, %
 which is given by the logarithm of the determinant of the quantum fields in the fixed background. The pioneer example, due to Schwinger~\cite{Schwinger:1951nm}, is of fermions in a constant background electric field. The example  we study in our paper is interesting, as we can find formulas for vacuum decay in generic field profiles (which satisfy a few technical assumptions that we state below).  Some exact results for generic field profiles were also obtained in \cite{Tomaras:2000ag,Tomaras:2001vs},  in $1+1$ dimensional QED.

Let us briefly review a case with no particle production. Consider free massless 
fermions interacting with a fixed non-Abelian gauge field background.
The effective action is obtained by the Gaussian integration over the fermion fields, and is given by a one loop determinant. 
If the field profile satisfies a ``good" behavior, that we specify later, the effective action is real and is expressed \cite{Polyakov:1984et} in terms of the Wess-Zumino-Novikov-Witten (WZNW) action \cite{Wess:1971yu,Novikov:1982ei,Witten:1983tw}. 
In this case particles are not created, since the vacuum decay rate is nonzero only when the effective action has an imaginary part.

Our goal is to determine the effective action for background fields that do lead to particle production. 
In this case, we have to discuss the in/out effective action which has an imaginary part, reflecting vacuum decay. The imaginary piece in the effective action is determined by a careful treatment of the Feynman $i\varepsilon$ prescription in a massless theory.

Our main result is that the effective action is modified by the inclusion of extra boundary terms, which are complex, and whose imaginary part gives the vacuum decay rate. The boundary term is a two-form which appears to be novel. To compute the boundary terms we need a certain Riemann-Hilbert decomposition. While the Abelian and non-Abelian decompositions are standard Riemann-Hilbert problems, the gravitational case has not been considered before.
The vacuum decay rate for Abelian background fields is given by the same formula of dissipative quantum mechanics obtained by Caldeira and Leggett~\cite{Caldeira:1981rx,Caldeira:1982uj}. Our results generalize their formulas for non-Abelian and gravitational backgrounds.

The rest of the paper is organized as follows. In section \ref{Absec} we compute the effective action and the new boundary term for an Abelian gauge field and 
discuss the general logic of the computation, which helps in the more complicated cases. In section \ref{nonAbsec} we find the effective action and the new boundary term for the non-Abelian gauge field. Finally, in section \ref{gravsec} we find the effective action and the new boundary term in the case of the gravitational field. In appendix \ref{bdyapp}, we discuss an alternative method of computation of the boundary terms. In appendix \ref{app:gauge-gravity}, we review the gauge-gravity duality between 
the non-Abelian and gravitational cases \cite{Alekseev:1988ce,Bershadsky:1989mf,Polyakov:1989dm}. Finally, in appendix \ref{CaldLeg}, we show the first perturbative correction to the Caldeira-Leggett formula coming from non-Abelian and gravitational backgrounds.

\section{Vacuum decay in an Abelian background}\label{Absec}
To set the stage, let us look 
at the Abelian case first. The  Lagrangian is 
\begin{align}
\mathcal{L} = \bar{\psi}  \gamma^{\mu} (i\partial_{\mu}+ A_{\mu})\psi\,=\bar{\psi}_-(i\partial_+ +A_+)\psi_- + \bar{\psi}_+ (i\partial_- +A_-)\psi_+ ,
\end{align}
where the metric is $\eta^{\mu\nu}=(1,-1)$, and we introduced light cone coordinates $x^{\pm}=(x^{0}\pm x^{1})/\sqrt{2}$. From the Lagrangian it is clear that the left movers $\psi_{+}$ and right movers $\psi_{-}$ are sourced by $A_-$ and $A_+$ fields, respectively. Therefore, the  determinant will split into a right-moving piece, a left-moving piece, and a contact term that ensures gauge invariance \cite{Polyakov:1984et}\footnote{In this section we write the effective action up to an unimportant overall factor $-\frac{e^{2}}{4\pi}$. In other words, we set $e^{2}=-4\pi$. The charge $e$ can be restored by the substitution $A_{\mu}\to eA_{\mu}$.}
\begin{align}\label{AbCompForm}
S(A_+,A_-) = \log \det (\gamma^{\mu}(i\partial_{\mu}+A_{\mu}))\,=W_+(A_+)+W_-(A_-)-2\int d^{2}x A_+A_-\, .
\end{align}
The contact term comes from short distance cutoff regulators; it is not related to particle production. In the case of strong fields which lead to particle production, $W_{+}$ and $W_{-}$ have imaginary parts.
The vacuum decay rate factorizes and is given by 
\begin{align}
|_{\textrm{out}}\langle 0 |0\rangle_{\textrm{in}}|^{2} = e^{-2 \textrm{Im}\, S(A)}=e^{-2\textrm{Im}W_+}e^{-2\textrm{Im}W_-}\,.
\end{align}
Let us compute the contribution to the effective action coming from $A_+$. We will treat $x^{+}$ as a time coordinate, while in the $x^{-}$ direction we assume that $A_+(x^{+},x^{-})\to0$ as $x^-\to\pm\infty$. An easy calculation of the diagram 
\begin{figure}[h!]
                \centering
                \includegraphics[width=6cm]{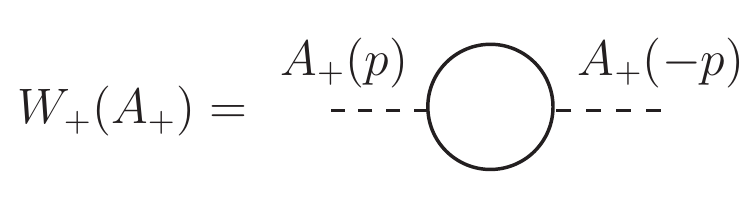}
\end{figure}

\noindent leads to ($d^{2}p=dp_{+}dp_{-}$)
\begin{align}\label{eamom}
W_+(A_{+}) =\int \frac{d^{2}p}{(2\pi)^{2}}\, \frac{p_{-}}{p_{+}+i\varepsilon\, \textrm{sgn}\,p_{-}} A_{+}(p)A_{+}(-p)\,.
\end{align}
As is well known, this result is exact and higher order corrections in $A_{+}$ are zero.  
The ``$i\varepsilon$" prescription follows from the Feynman rule $\frac{1}{p^{2}} \Rightarrow \frac{1}{p^{2}+i\varepsilon}=\frac{1}{p_{-}}\big(\frac{1}{p_{+}+i\varepsilon\, \textrm{sgn}\,p_{-}}\big)$. The term in parenthesis is the 
Feynman Green's function. We see that\footnote{We use $1/(p_{+}+i\varepsilon \textrm{sgn}\, p_{-})= \mathcal{P}(1/p_{+})-i \pi\,  \textrm{sgn}\,( p_{-}) \delta(p_{+}) $. } 
\begin{align}
\textrm{Im} \,W_+(A_{+})  =-\int \frac{d^{2}p}{4\pi} |p_{-}| \delta(p_{+})A_{+}(p)A_{+}(-p)  =-\frac{1}{4\pi}\int dp_{-}|p_{-}| A_{+}(0,p_{-})A_{+}(0,-p_{-})\,. \label{EffInPspace}
\end{align}
The condition of vacuum stability ($\textrm{Im}\,W_+=0$) is thus $\int_{-\infty}^{+\infty} A_{+}(y^{+},x^{-})dy^{+}=0$. It is useful to rewrite the formula (\ref{EffInPspace}) in position space. If we denote 
\begin{align}
\omega(x^{-})\equiv \int_{-\infty}^{+\infty}A_{+}(y^{+},x^{-})dy^{+}\,, \label{abomega}
\end{align}
from (\ref{EffInPspace}) we obtain\footnote{The fact that the effective action depends on $\omega(x^{-})$ demonstrates that the gauge symmetry  in our system is restricted by the condition that gauge transformations  for $A_{+}$ and $A_-$ must be trivial at the boundary of spacetime.} 
\begin{align}
\textrm{Im}\, W_+(A_{+}) = \frac{1}{4\pi}\int_{-\infty}^{+\infty} dx^{-}dy^{-} \frac{(\omega(x^{-})-\omega(y^{-}))^{2}}{(x^{-}-y^{-})^{2}}\,. \label{imw1}
\end{align}
We recognize this formula as the friction term in Caldeira-Leggett's dissipative quantum mechanics \cite{Caldeira:1981rx,Caldeira:1982uj}. Below we will find the non-Abelian and gravitational generalizations of this action.

 It is instructive to rewrite (\ref{imw1}) in a slightly different form. Let us introduce two complex functions $\omega_{\textrm{up}}(x^{-})$ and $\omega_{\textrm{down}}(x^{-})$, which are analytic 
in the upper and lower half-planes, respectively. They are related to $\omega(x^{-})$ as 
\begin{align}\label{RHScalar}
 \omega_{\textrm{up}}(x^{-})-\omega_{\textrm{down}}(x^{-})=\omega(x^{-})
\end{align}
for real $x^-$. This decomposition of the function $\omega(x^-)$ is called the {\it scalar Riemann-Hilbert problem} and the explicit solution in this case is given by
\begin{align}
\omega_{\textrm{up}/\textrm{down}}(x^{-}) = \frac{1}{2\pi i}\int_{-\infty}^{+\infty} \frac{\omega(y^{-})dy^{-}}{y^{-}-x^{-}\mp i \varepsilon}\,. \label{omupdown}
\end{align}
In terms of $\omega_{\textrm {up/down}}$, the imaginary part of the effective action can be written as 
 \begin{align}\label{EffAcRH}
\boxed{\textrm{Im}\, W_+(A_{+}) = \textrm{Im}\int_{-\infty}^{+\infty}dx^{-} \left(\omega_{\textrm{down}}\partial_- \omega_{\textrm{up}} \right)}\,.
\end{align}
The generalization of the formula (\ref{EffAcRH}) for the strong non-Abelian and gravitational cases is the main goal of this paper. 

There is yet another way of obtaining (\ref{EffAcRH}), which will be useful below. We can parametrize $A_{+}$ as  
 \begin{align}
A_+(x^+,x^-)=\partial_+\phi(x^+,x^-)\,
\end{align}
and we notice that the ``Wilson line" $\phi(x^+,x^-)$ has residual gauge invariance $\phi \to \phi + u(x^-)$. We say that  $\phi =\phi_R(x^+,x^-)$  is in {\it retarded} gauge if it obeys the boundary condition $\phi_{R}(-\infty,x^-)\to0$ and therefore
\begin{align}
\phi_{R}(x^+,x^-)= \int_{-\infty}^{x^+}A_+(y^{+},x^-)dy^{+} \,.
\end{align}
We see that $\phi_R$ is manifestly real and causal, as $\phi_R(x^+,x^-)$ only depends on $A_+(y^+,x^-)$ for $y^+<x^+$; moreover, $\phi_R(+\infty,x^-)=\omega(x^-)$, so the imaginary part of the effective action (\ref{EffAcRH}) is written in terms of the boundary value of $\phi_R$ and the whole $W_{+}(A_{+})$ reads
\begin{align}
W_{+}(A_{+}) = \int d^{2}x \, \partial_{+}\phi_{R}\partial_{-}\phi_{R}+\int_{-\infty}^{+\infty}dx^{-} \left(\, \omega_{\textrm{down}}\partial_- \omega_{\textrm{up}} \right)\,. \label{abeffret}
\end{align}
We can use the solution of the Riemann-Hilbert problem (\ref{RHScalar}) and the residual gauge invariance of $\phi$ to define a {\it spectral} (or Feynman) gauge, namely
\begin{align}\label{spec gauge}
\phi_S(x^+,x^-)\equiv\phi_R(x^+,x^-)-\omega_{\textrm{down}}(x^-)\to\begin{cases} \omega_{\textrm{up}}(x^-), &x^+\to+\infty\\  -\omega_{\textrm{down}}(x^-), &x^+\to-\infty\end{cases}.
\end{align}
In the spectral gauge  the effective action (\ref{abeffret}) reads
\begin{align}
W_{+}(A_{+}) = \int d^{2}x \,\, \partial_{+}\phi_{S}\partial_{-}\phi_{S}\,, \label{abeffspec}
\end{align}
and has the form of the usual result~\cite{Schwinger:1962tp}. In our case, the difference is that the function $\phi_{S}$ is complex valued and  (\ref{abeffspec}) contains both real and imaginary parts of the effective action! The conclusion is that in  the {\it spectral}  gauge, we do not require boundary terms in the effective action, whereas in the  {\it retarded} gauge, we  have boundary terms, which are complex and account for the vacuum decay.

The logic is summarized as follows. If we use the {\it spectral} gauge, 
then the expressions for the effective actions are well known \cite{Schwinger:1962tp, Polyakov:1984et, Polyakov:1987zb}, as the boundary terms evaluate to zero. Then passing from the {\it spectral} gauge to {\it retarded}  gauge we determine the functional form of the boundary terms. In the retarded gauge, the boundary term
contains the imaginary part of the effective action.  In Appendix \ref{bdyapp} we discuss an alternative method to compute the full effective action, by exploiting (chiral or trace) anomaly equations.

\section{Vacuum decay in a non-Abelian background}\label{nonAbsec}
In the non-Abelian case the general form of the effective action reads
\begin{align}\label{nAbCompForm}
S(A_+,A_-)= \log \det (\gamma^{\mu}(i\partial_{\mu}+A_{\mu}))=W_+(A_+)+W_-(A_-)+2\int\, d^{2}x\, \Tr (A_+A_-)\,
\end{align}
and  imaginary terms responsible for the particle production are present only in $W_{+}$ and $W_{-}$. We concentrate again on $W_{+}(A_{+})$, which is formally given by the following sum of Feynman diagrams
\begin{figure}[h!]
                \centering
                \includegraphics[width=9.5cm]{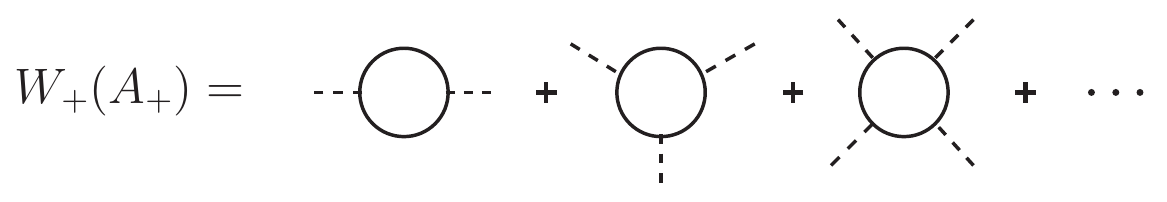}
\end{figure}

\noindent If we parametrize $A_{+}=g^{-1}\partial_{+}g$  we get $W_{+}(g)$. If $g(x^{+}\to \pm \infty,x^{-})=\mathbbm{1}$ then $W_{+}(g)$ is the WZNW action~\cite{Polyakov:1983tt}
\begin{align}\label{wznwact}
W_{\rm WZNW}(g)\equiv {1 \over 2} \int \dd^{2}x\, \Tr (\partial^{\mu}g^{-1}\partial_{\mu}g) - {1 \over 3}\int \dd^{2}x dt \,\varepsilon^{\mu\nu\lambda} \Tr (g^{-1}\partial_{\mu}{g} g^{-1}\partial_{\nu}g g^{-1}\partial_{\lambda}g)\,,
\end{align}
where  in the last Wess-Zumino (WZ) term we introduced the extra $t$-dependence: $g(x^{+},x^{-},t)$ such that $g(x^{+},x^{-},0)=\mathbbm{1}$ and $g(x^{+},x^{-},1)= g(x^{+},x^{-})$; and $\mu,\nu,\lambda = (\pm,0)$, and $\varepsilon^{0-+}=1$, where zero corresponds to the $t$ coordinate.

From the Abelian case, we expect that vacuum decay occurs for $A_{+}$ with  $g^{-1}(-\infty,x^{-})$ $\cdot g(+\infty,x^{-})\neq \mathbbm{1}$, or, in different notation:
\begin{align}
\Omega(x^{-}) \equiv P \exp \int_{-\infty}^{\infty} A_{+}(y^{+},x^{-}) dy^{+}\neq \mathbbm{1},
\end{align}
where ``$P \exp$" is the path-ordered exponential.
In this case the effective action is not given by (\ref{wznwact}); it must include new boundary terms. Indeed, looking at the variation of the WZ term
 \begin{align}
S_{\textrm{WZ}} \equiv \int d^{2}x dt \,  \varepsilon^{\mu\nu\lambda} \textrm{Tr}(a_{\mu}a_{\nu}a_{\lambda})\,, 
\end{align}
where  $a_{\mu}\equiv g^{-1} \partial_{\mu}g$, with $\delta a_{\mu}= \nabla_{\mu}\varepsilon$, we obtain
\begin{align}
\delta S_{\textrm{WZ}} &= \int d^{2}x dt \, \varepsilon^{\mu\nu\lambda} \textrm{Tr}(a_{\mu}a_{\nu}\nabla_{\lambda}\varepsilon) \sim \int d^{2}x dt \, \varepsilon^{\mu\nu\lambda} \nabla_{\lambda}\textrm{Tr}\big( (\partial_{\mu}a_{\nu}-\partial_{\nu}a_{\mu})\varepsilon\big)  \notag\\
&=\int d^{2}x dt \,\varepsilon^{ij}\,\partial_{0} \textrm{Tr}(\partial_{i}a_{j} \varepsilon)  - \int d^{2}x dt \, \partial_{+} \textrm{Tr} \big((\partial_{-}a_{0}-\partial_{0}a_{-})\varepsilon\big) \notag\\
&= \int d^{2}x \, \varepsilon^{ij} \textrm{Tr} (\partial_{i}a_{j}\varepsilon) -\int dx^{-} dt\, \left.\textrm{Tr}\big((\partial_{-}a_{0}-\partial_{0}a_{-})\varepsilon\big)\right|^{x^{+}=+\infty}_{x^{+}=-\infty}\,. \label{varWZ}
\end{align}
The first term here is standard while the time-boundary term explicitly violates $t$-symmetry. In other words, $S_{\textrm{WZ}}$ is dependent on the $t$-parametrization.  This unphysical dependence on the extrapolation disappears when the right boundary terms are added to WZNW action.

Notice that the matrix $g$ has a gauge symmetry 
 \begin{align}
g(x^{+},x^{-})\to u(x^{-})g(x^{+},x^{-}) \,,
\end{align}
where $u(x^{-})$ is an arbitrary complex matrix. The retarded gauge is defined by
\begin{align}
g_R(x^{+} \to -\infty, x^{-})= \mathbbm{1}\ \Rightarrow g_{R}(x^{+},x^{-}) = P \exp \int_{-\infty}^{x^{+}} dy^{+} A_{+}(y^{+},x^{-})\, .
\end{align} 
Like in the Abelian case, we see that $\Omega(x^-)=g_R(+\infty,x^-)$. 

Proceeding by analogy, we should look for complex valued matrices $\Omega_{\textrm{down}}(x^{-})$ and $\Omega_{\textrm{up}}(x^{-})$ that are a solution to the {\it matrix Riemann-Hilbert problem}
 \begin{align}\label{matriemhilb}
 \Omega_{\textrm{down}}(x^{-}) \Omega_{\textrm{up}}(x^{-})=\Omega(x^{-}) \,, 
\end{align}
for real values of $x^-$. We assume that  $\Omega^{-1}_{\textrm{up}}(x^{-})$ and $\Omega^{-1}_{\textrm{down}}(x^{-})$ are also analytic in the upper and lower half-planes, respectively. Unfortunately, the matrix Riemann-Hilbert problem does not have an explicit general solution.\footnote{For a review on the subject and the cases where an explicit solution is available, see \cite{gohberg2003overview}. Notice that the right and left decompositions are inequivalent, namely, we could look for $\Omega(x^-)=\widetilde{\Omega}_{\textrm{up}}(x^{-}) \widetilde{\Omega}_{\textrm{down}}(x^{-})$, but in terms of these matrices we do not obtain spectral boundary conditions in a simple way. In general $\widetilde{\Omega}_{\textrm{up}/\textrm{down}}(x^{-})\ne{\Omega}_{\textrm{up}/\textrm{down}}(x^{-})$. We thank A. Kisil for discussions on the matrix Riemann-Hilbert problem.}

As we see from (\ref{varWZ}) the retarded gauge choice requires extra terms in the WZ term in order to cancel the unacceptable boundary contributions. However, we can use our gauge freedom in choosing $g$ to eliminate the boundary terms. Let us introduce the spectral (or Feynman) gauge: 
 \begin{align}
g_{S}(x^{+},x^{-}) \equiv \Omega^{-1}_{\textrm{down}}(x^{-})\ g_{R}(x^{+},x^{-}) \to\begin{cases} \Omega_{\textrm{up}}(x^-),&x^+\to+\infty\\  \Omega^{-1}_{\textrm{down}}(x^-),&x^+\to-\infty \end{cases}. \label{feyngauge}
\end{align}
It follows from here that $g_{S}(x^{+},x^{-})$ at $x^{+}\to \pm \infty$ is analytic in the lower/upper half-planes and thus all boundary terms vanish after $x^{-}$ integration. By analogy with the Abelian case, we come to the conclusion that in the spectral gauge there are no boundary terms! The effective action is just the standard WZNW action (\ref{wznwact}), which is complex valued, as $g_S$ is complex 
 \begin{align}\label{easg}
W_{+}(A_{+})= W_{\rm WZNW}(g_{S})\,.
\end{align}
A more physical justification of the absence of boundary terms in the spectral gauge is discussed in the Appendix \ref{bdyapp}. 

From (\ref{easg}), we now determine the boundary term that must be present in the effective action written in an arbitrary gauge. For example, in going from spectral to retarded gauge, we do not change $A_{+}=g^{-1}\partial_{+}g$, therefore the effective actions must be the same,
\beq\label{releffa}
W_{+}(A_{+})=W_{\rm WZNW}(\Omega^{-1}_{\textrm{down}} g_{R})=W_{\rm WZNW}(g_{R})+W_B(\Omega_{\rm up},\Omega_{\rm down})\,.
\eeq

\begin{figure}[h!]
                \centering
                \includegraphics[width=3.9cm]{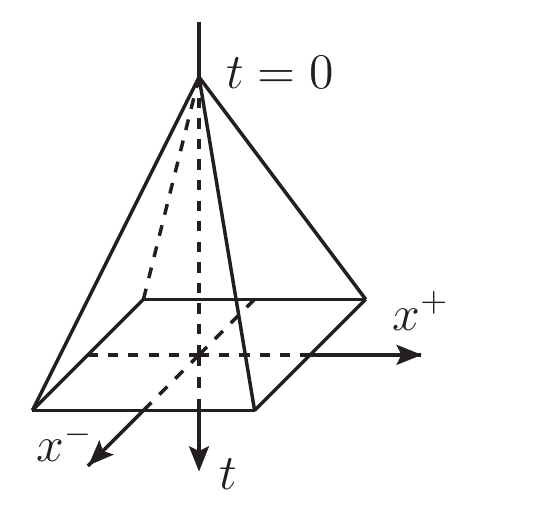}
                \caption{Penrose-Nefertiti diagram. The usual Penrose diagram of $1+1$ dimensional Minkowski spacetime is the base of a pyramid. The embedding coordinate $t$ runs from the apex ($t=0$) to the base ($t=1$). The new boundary terms in the effective action are supported at the $t$ - $x$ faces of the pyramid.}
                \label{Nefertiti}
\end{figure}

In order to proceed we use exterior calculus to derive a composition formula for the WZ term. Let us introduce two $1$-forms $a$ and $b$, with $a= g^{-1} d g, b = dh h^{-1}\,$. $a$ and $b$ satisfy the equations $da = -a\wedge a$, $d b= b\wedge b\,$. 
Consider the $1$-form $c = (gh)^{-1}d(gh) = h^{-1}(a+b)h$. Then we have  
\begin{align}
\Tr (c\wedge c \wedge c)&=\Tr(a \wedge a\wedge a)+\Tr(b\wedge b\wedge b) -3d (\Tr\, a \wedge b)\,. \label{rel}
\end{align}
Now we apply (\ref{rel}) with $g_S=\Omega_{\rm down}^{-1}g_R$. From the quadratic term in the WZNW action we obtain\footnote{In light-cone coordinates $\Tr(\partial^{\mu}g^{-1}\partial_{\mu}g) =\Tr(\partial_{-}g^{-1}\partial_{+}g)+\Tr(\partial_{+}g^{-1}\partial_{-}g)=2\Tr(\partial_{+}g^{-1}\partial_{-}g)$.} 
\beq\label{chh}
{1 \over 2} \int \dd^{2}x\, \Tr (\partial^{\mu}g_{S}^{-1}\partial_{\mu}g_{S}) =\frac{1}{2}\int \dd^{2}x\, \Tr (\partial^{\mu}g_{R}^{-1}\partial_{\mu}g_{R})\,+\int \dd^{2}x \, \Tr (\partial_-\Omega_{\rm down} \Omega_{\rm down}^{-1}\partial_{+}g_{R} g_{R}^{-1})\,,
\eeq
and using (\ref{releffa}) and (\ref{rel}) for the WZ term in (\ref{wznwact}) we find
\begin{align}\label{wzm1}
W_{\textrm{WZ}}(g_{S})=W_{\textrm{WZ}}(g_{R})-  3\int_{(x^{+},x^{-},t)}  d\,\big( \Tr (\Omega_{\rm down} d \Omega^{-1}_{\rm down} \wedge dg_{R}g_{R}^{-1} )\big)\,. 
\end{align}
Notice that the Penrose diagram for our space-time with the embedding dimension is a pyramid; we call it the Penrose-Nefertiti diagram (see figure \ref{Nefertiti}). 

\noindent The first term in (\ref{wzm1}) is real (we assume $A_+$ is real), while the boundary term, which has support at the faces of the pyramid, is complex valued.    Using Stokes' theorem in (\ref{wzm1}), we obtain
\begin{align}
W_{\textrm{WZ}}(g_{S})=W_{\textrm{WZ}}(g_{R})-  3\int d^{2}x  \Tr (\Omega_{\rm down} \partial_{-} \Omega^{-1}_{\rm down} \partial_{+}g_{R}g_{R}^{-1} )+  3\int\limits_{(x^{-},t)}   \Tr (\Omega_{\rm down}^{-1} d \Omega_{\rm down} \wedge \Omega_{\rm up} d\Omega_{\rm up}^{-1})\,. \label{wzbound}
\end{align}
Using (\ref{releffa}), (\ref{chh}) and (\ref{wzbound}) we finally obtain 
\begin{align}
\boxed{W_{B}(\Omega_{\rm up},\Omega_{\rm down})= \int\limits_{(x^{-},t)}   \Tr (\Omega_{\rm down}^{-1} d \Omega_{\rm down} \wedge \Omega_{\rm up}d\Omega_{\rm up}^{-1})\,}\,\,. \label{nonAbBound}
\end{align}
The formula (\ref{nonAbBound}) is one of the main results of our paper.\footnote{The effective action for arbitrary $g$ is $W_{+}(A_{+}) = W_{\rm WZNW}(g^{-1}(-\infty,x^{-})g)+W_{B}(\Omega_{\rm up},\Omega_{\rm down})$.} The boundary term is complex valued, and although not manifestly imaginary, contains the imaginary part of the effective action\footnote{Notice that, although the boundary term does depend on the $t$-interpolation, its imaginary part does not! One can see this by looking at the variation of (\ref{nonAbBound}), $
\delta W_{B}=\frac{1}{2} \int dx^{-} \Tr ( \Omega_{\rm down}^{-1}\delta \Omega_{\rm down}\Omega_{\rm up}\partial_{-}\Omega_{\rm up}^{-1}+ \Omega_{\rm down}^{-1}\partial_{-}\Omega_{\rm down} \delta \Omega_{\rm up}\Omega_{\rm up}^{-1})+\frac{1}{2}\int dt dx^{-}\Tr([\Omega^{-1}\partial_{0}\Omega,\Omega^{-1}\partial_{-}\Omega]\Omega^{-1}\delta \Omega) \, .$
We see that the last term is $t$-dependent but explicitly real ($\Omega$ is a real matrix), whereas the first term is $t$-independent and complex. We also notice that the $t$-dependent term cancels with the $t$-dependent term in the variation of the WZ term (\ref{varWZ}) in the effective action. Thus the variation of the effective action is also $t$-independent. }. 

We emphasize that this boundary term is the non-Abelian generalization of the Caldeira-Leggett dissipative term, and is given by a  two-form\footnote{A similar two-form was found in Euclidean manifolds with a boundary, in~\cite{Baumgartl:2004iy, Baumgartl:2006xb}. We thank N. Nekrasov for bringing these papers to our attention.}. We also notice  that our two-form is a Minkowski space counterpart of the  Atiyah-Patodi-Singer $\eta$-invariant \cite{Atiyah:1975jf,Atiyah:1976jg,Atiyah:1980jh}, which appears in  Euclidean manifolds with boundary. We present the leading order, non-Abelian correction to the Caldeira-Leggett formula in Appendix \ref{CaldLeg}.

\section{Vacuum decay in the gravitational field}\label{gravsec}
Now we consider a theory of fermions coupled to a fixed gravitational field.  It is convenient to parametrize the metric in the light cone coordinates,
\beq\label{mlcg}
ds^2=h_{+-}(x^+,x^-)dx^+dx^-+h_{++}(x^+,x^-)dx^+dx^++h_{--}(x^+,x^-)dx^-dx^- \,.
\eeq
We assume that the background fields are asymptotically flat, i.e. $h_{++}(x^{+},x^{-})\to0$, $h_{--}(x^{+},x^{-})\to0$ and $h_{+-}(x^+,x^-)\to 1$ as $x^\pm\to\pm\infty$.
The Lagrangian is\footnote{For simplicity, we consider Majorana fermions. As in \cite{Knizhnik:1988ak}, we perform a field redefinition to write the Lagrangian in the form (\ref{lgra}). In the previous sections, we considered Dirac fermions, as these can carry electric and color charge.}
\beq\label{lgra}
{\cal L} =  \psi_-(\partial_+-h_{++}\partial_-)\psi_-+\psi_+(\partial_- -h_{--}\partial_+)\psi_+\,.
\eeq
Like in the non-Abelian case, the effective action is 
 \begin{align}
S(h_{++},h_{--},h_{+-})=W_{+}(h_{++})+W_{-}(h_{--})+L(h_{++},h_{--},h_{+-})\,,
\end{align}
where the last term $L$ is a  local and real term and appears due to the UV regulator. We concentrate on the calculation of the contribution from left-moving fermions, $W_+(h_{++})$.
For gravity we use the same logic as in the non-Abelian case.  We parametrize the metric tensor $h_{++}(x^{+},x^{-})$ using the function $f(x^{+},x^{-})$ defined by the equation
\begin{align}
(\partial_{+}-h_{++}\partial_{-})f=0\,, \label{defoff}
\end{align} 
which is a gravitational analog of the Wilson line. Lines of constant $f$ correspond to the characteristics of light-like, right-moving geodesics in the background spacetime. Notice that there is an ambiguity in $f$, namely 
\begin{align}
f(x^{+},x^{-})\; \Rightarrow \; u(f(x^{+},x^{-}))\,, \label{gaugesymgrav}
\end{align}
where $u(x^{-})$ is an arbitrary invertible complex function of one variable. 
The {\it retarded} gauge is defined by 
\beq\label{ordgbc}
f_{R}(x^{+}\to -\infty,x^{-})=x^-\Rightarrow f_{R}(x^{+},x^{-}) \equiv P \exp \Big( \int_{-\infty}^{x^{+}} dy^{+}h_{++}(y^{+},x^{-})\partial_{-}\Big) x^{-}\,.
\eeq
As in the case of gauge fields, we need to add a suitable boundary term to the effective action~\cite{Polyakov:1987zb}
\begin{align}
W_{\textrm{gWZ}}(f)\equiv \int \dd^{2}x \left({\partial_{-}^{2}f \partial_{+}\partial_{-}f \over (\partial_{-}f)^{2}}-{(\partial_{-}^{2}f)^{2}\partial_{+}f \over (\partial_{-}f)^{3}}\right)\,, \label{la}
\end{align}
where gWZ stands for ``gravitational Wess-Zumino" and we omit an overall normalization factor, which is $-1/48\pi$ in our case. Alternatively, we can use the gauge symmetry (\ref{gaugesymgrav}) to eliminate the boundary term. Let us introduce the {\it spectral} gauge by 
\beq\label{spgg}
f_{S}(x^{+},x^{-})\equiv \Gamma^{-1}_{\rm down}(f_{R}(x^{+},x^{-}))=\begin{cases}\Gamma_{\rm up}(x^{-}), \,~~~~~x^{+}\to +\infty\,, \cr
 \Gamma^{-1}_{\rm down}(x^{-}), ~~~x^{+}\to -\infty\,, \end{cases}
\eeq
where $\Gamma_{\rm up}(x^{-})$ and $\Gamma_{\rm down}(x^{-})$ are analytic functions in the upper and lower $x^-$ half-plane. We also assume that the inverse functions $\Gamma_{\rm up}^{-1}(x^{-})$ and $\Gamma^{-1}_{\rm down}(x^{-})$ are analytic in the upper and lower $x^-$ half-plane respectively. In this case, to determine $\Gamma_{\rm up, down}$, we need to solve a {\it ``functional'' Riemann-Hilbert problem} \footnote{In an analogous fashion to the matrix Riemann-Hilbert problem, we can have right and left decompositions of the function $f$. Namely, we can consider functions $\tilde{\Gamma}_{\rm up/down}$ such that $ \tilde{\Gamma}_{\rm up}(\tilde{\Gamma}_{\rm down}(x^{-}))=\Gamma(x^{-}) $ in the real line. In general, $\tilde{\Gamma}_{\rm up/down}\ne \Gamma_{\rm up/down}$.},
\beq\label{frhp}
\Gamma_{\rm down}(\Gamma_{\rm up}(x^{-})) =\Gamma(x^{-}) \,,
\eeq
where 
\begin{align}
\Gamma(x^{-}) \equiv P \exp \Big( \int_{-\infty}^{+\infty} dy^{+}h_{++}(y^{+},x^{-})\partial_{-}\Big) x^{-} = f_{R}(+\infty,x^{-})\,.\label{gammagrav}
\end{align}
To our knowledge, the Riemann-Hilbert problem (\ref{frhp}) has not been considered in the mathematics literature before. We also notice that (\ref{frhp}) doesn't  have an explicit solution. \footnote{Finding a physically relevant explicit solution to (\ref{frhp}) seems to be hard. On the other hand one can find solutions in terms of meromorphic funtions. For example, $\Gamma_{\rm down}(x)= \frac{\epsilon}{1-x_{3}^{2}}\frac{(x-a)^{2}}{(x-ix_{1})(x-i x_{2})}$, $\Gamma_{\rm up}(x)= \frac{ax-b}{x+ix_{3}}$ and $\Gamma(x)=\frac{\epsilon}{1+x^{2}}$ is a solution to (\ref{frhp}), where $a=\frac{i}{2}(x_{1}+x_{2}+(x_{1}-x_{2})x_{3})$, $b= \frac{1}{2}(x_{1}-x_{2}+(x_{1}+x_{2})x_{3})$, $x_{1},x_{2},x_{3}>0$ and $\epsilon$ is an arbitrary real parameter.}

By similar arguments as in the previous sections, the effective action is 
\begin{align}
W_{+}(h_{++})=W_{\textrm{gWZ}}(f_{S})\,, \label{effgravspect}
\end{align}
where $W_{\textrm{gWZ}}$ is given by (\ref{la}). See also Appendix \ref{bdyapp} for a different derivation of (\ref{effgravspect}).
This effective action is complex valued. In retarded and spectral gauges the metric $h_{++}(x^{+},x^{-})$ is the same, therefore we have the equality
\beq\label{effrelg}
W_{+}(h_{++}) = W_{\rm gWZ}(\Gamma_{\rm down}^{-1}(f_{R})) =W_{\rm gWZ}(f_{R})+W_{B}(\Gamma_{\rm up},\Gamma_{\rm down})\,.
\eeq
Using (\ref{la}), (\ref{effrelg})  we get
\beq\label{gbac}
W_{B}(\Gamma_{\rm up},\Gamma_{\rm down})=\int d^{2}x \,\partial_{-}f_{R}\, {(\Gamma_{\rm down}^{-1})'' \over (\Gamma_{\rm down}^{-1})'}\,\partial_{-}\left({\partial_{+}f_{R} \over \partial_{-}f_{R}}\right).
\eeq
Now, introducing new variables $y^- = f_{R}(x^{+},x^{-})$, $y^{+}=x^{+}$  one can get\footnote{To arrive at this formula we need two steps. At step 1 we define the inverse function $f_{R}^{-1}(\cdot ,\cdot)$ by $
f_{R}^{-1}(x^{+},f_{R}(x^{+},x^{-}))=x^{-}$ and notice that $\int d^{2}x \partial_{-}f_{R}=\int d^{2}y$ and $\partial_{-}=(\partial_{-}f_{R})\partial/\partial y^{-}=(\partial f_{R}^{-1}/\partial y^{-})^{-1}\partial/\partial y^{-}$ and  $\partial_{+}f_{R}/\partial_{-}f_{R}=-\partial f_{R}^{-1}/\partial y^{+}$. At step 2 we integrate over $y^{+}$  and use that $\partial f_{R}^{-1}/\partial y^{-}(+\infty,y^{-}) =1/\Gamma'(\Gamma^{-1}(y^{-}))$ and $ \partial f_{R}^{-1}/\partial y^{-}(-\infty,y^{-}) =1$. It was crucial here to assume that $f_{R}(x^{+},x^{-})$ is invertible for all $x^{+}$. }  
\begin{align}
W_{B}(\Gamma_{\rm up},\Gamma_{\rm down})=\int dy^{-}  {\partial \over \partial y^{-}} \log \big((\Gamma^{-1}_{\rm down})'(y^{-})\big)  \log \big(\Gamma'(\Gamma^{-1}(y^{-}))\big). \label{gbaccc}
\end{align}
Finally introducing a coordinate $s = \Gamma^{-1}(y^{-})$ and using that $(\Gamma^{-1}_{\rm down})'(\Gamma(s)) = 1/\Gamma'_{\rm down}(\Gamma_{\rm up}(s))$ and $\Gamma'(s)= \Gamma'_{\rm down}(\Gamma_{\rm up}(s))\Gamma'_{\rm up}(s)$ we obtain\footnote{The expression (\ref{gbaccc}) is very similar to formula (5.23) in \cite{Chung:1993rf}. The reason for the similarity of the results is puzzling to us and is an interesting open question. We thank H. Verlinde for pointing this to us.}  
\begin{align}
\boxed{W_{B}(\Gamma_{\rm up},\Gamma_{\rm down})=\int  \dd s \log \big( \Gamma'_{\rm down}(\Gamma_{\rm up}(s))\big){\partial \over \partial s}   \log \big( \Gamma'_{\rm up}(s)\big)\,.} \label{gbacccc}
\end{align}
Therefore the effective action in the retarded gauge is 
\beq\label{gravefa}
W_{+}(h_{++}) =   \int \dd^{2}x \left({\partial_{-}^{2}f_{R} \partial_{+ -}f_{R} \over (\partial_{-}f_{R})^{2}}-{(\partial_{-}^{2}f_{R})^{2}\partial_{+}f_{R} \over (\partial_{-}f_{R})^{3}}\right)  + \int  \dd s \log \big( \Gamma'_{\rm down}(\Gamma_{\rm up}(s))\big){\partial \over \partial s}   \log \big( \Gamma'_{\rm up}(s)\big).
\eeq
The bulk term is manifestly real, while the boundary term is complex, and, in particular, contains the imaginary piece of the effective action. 
 
In appendix \ref{app:gauge-gravity}, we review a connection between gauge theory and gravity in two dimensions~\cite{Alekseev:1988ce,Bershadsky:1989mf,Polyakov:1989dm}, and phrase \eqref{frhp} in terms of a matrix Riemann-Hilbert problem, in the hope that this simple connection might be useful in finding explicit solutions of the functional Riemann-Hilbert problem.

\bigskip
\section{Conclusions}\label{sec:conclusions}
We conclude with a few open questions that we find interesting:
\begin{itemize}
\item We considered fixed background fields. One can also integrate over these backgrounds, in a similar fashion as in perturbative string theory. Does the boundary term play any role in that case?
\item Is the functional Riemann-Hilbert problem solvable for some set of functions? Perhaps there are relevant gravitational backgrounds for which one could compute the vacuum decay rate explicitly. 
\item It would be quite interesting to classify the backgrounds that, although curved, keep the vacuum stable.
\item The boundary terms in the non-Abelian and gravitational cases are complex valued. We tried, but could not find a compact expression for the imaginary part of the effective action. In particular, this supposed expression for the imaginary part, in the non-Abelian case, should be manifestly $t$-independent.
\item If we use retarded Wilson lines, then we can write causal equations of motion for the background fields which include quantum friction. Can the quantum friction screen certain backgrounds once we solve for them dynamically?

\item In the gravitational case we assume that $f_{R}(x^{+},x^{-})$ is invertible for any $x^{+}$. An interesting variation of this property is the case when $f_{R}(x^{+}=+\infty,x^{-})$ is not invertible. This loss of information may be related to the backgrounds with horizons, which have intrinsic entropy.
\end{itemize}
In summary, we are still scratching the surface in terms of potential applications of these new results.
\subsubsection*{Acknowledgements}

We thank T. Banks, D. Baumann, J. Cardy, G. Dunne,  R. Flauger, A. Kisil, I. Klebanov, C. Mafra, N. Nekrasov, H. Osborn,  M. Rangamani, H. Reall, S. Shenker, H. Verlinde and A. Zhiboedov for helpful discussions. We also thank D. Baumann for comments on a draft. 
G.L.P.~thanks the Aspen Center for Physics (supported in part by NSF Grant PHY10-66293) and the University of Amsterdam for their hospitality. He also thanks the KITP for hospitality during the program `Quantum Gravity Foundations: UV to IR'. Research at the KITP is supported in part by the National Science Foundation under Grant No. NSF PHY11-25915. G.L.P.~acknowledges support from a Starting Grant of the European Research Council (ERC STG grant 279617). The work of A.M.P. and G.M.T. was supported in part by the US NSF under Grant No. PHY-1314198.

\appendix

\bigskip
\section{Boundary conditions on induced currents and alternative derivation of the boundary actions}\label{bdyapp}

In this appendix we derive the effective action for non-Abelian and gravitational cases using the anomaly equations.
We start with the non-Abelian case.
We define $J_{\mu}\equiv\delta W/\delta A_{\mu}$; then the anomaly equations read \cite{Polyakov:1983tt}\footnote{To restore the unimportant overall factor in front of  the effective action one needs to replace $\varepsilon^{\mu\nu} F_{\mu\nu} \to \frac{\varepsilon^{\mu\nu}}{2\pi} F_{\mu\nu}$. }
\begin{align}
\begin{cases}
\partial_{\mu}J^{\mu}+[A_{\mu},J^{\mu}]=0\,, \\
\varepsilon^{\mu\nu}(\partial_{\mu}J_{\nu}+[A_{\mu},J_{\nu}])= \varepsilon^{\mu\nu} F_{\mu\nu}\,.
\end{cases}
\end{align}
Working in the light-cone cone coordinates $x^{\pm}$ and choosing the axial gauge $A_{-}=0$, we get $\partial_{-}(A_{+}-J_{+})=0$  and ($\varepsilon^{-+}=1$)
\begin{align}
\partial_{-}A_{+}+\partial_{+}J_{-} -[J_{-},A_{+}]=0. \label{anomeq1}
\end{align}
Parametrizing $A_{+}=g^{-1}\partial_{+}g$ one can find that the general solution of  (\ref{anomeq1}) is
\begin{align}
J_{-}=-g^{-1}\partial_{-}g-g^{-1}j_{-}g\,,
\end{align}
where $j_{-}=j_{-}(x^{-})$ is, at this stage, an arbitrary complex matrix function, which depends only on $x^{-}$, and has to be fixed by additional physical arguments. On the other hand the variation of the effective action is 
\begin{align}
\delta W(A_{+}) = \int d^{2}x \, \Tr (J_{-}\delta A_{+})\,. \label{varofw}
\end{align}
As we will see below, it is exactly the term $g^{-1}j_{-}g$ in the current $J_{-}$ which is responsible for the imaginary part of the effective action. 

In order to fix  $j_{-}(x^{-})$ we use the ``analyticity" argument. Namely we say that the 
induced current ${}_{\rm out}\langle J_-(x^+,x^-) \rangle_{\rm in}$  must satisfy the analytical (spectral) boundary conditions\footnote{Although $J_{-}(x^{+},x^{-})$ is a hermitian operator, the matrix element ${}_{\rm out}\langle J_-(x^+,x^-) \rangle_{\rm in}$ can be complex valued, as we are not computing an expectation value of the current for a given state, but rather evaluating a transition amplitude between states without particles in the past and without particles in the future.
}:
\beq\label{bc}
{}_{\rm out}\langle J_-(x^+,x^-) \rangle_{\rm in}\to \begin{cases}J_{\rm up}(x^-),& x^+\to+\infty\,,\\ J_{\rm down}(x^-),& x^+ \to -\infty\,,\end{cases}
\eeq
where $J_{\rm up}(x^{-})$ and $J_{\rm down}(x^{-})$ are complex matrix functions analytic in the upper and lower $x^-$ half-planes correspondingly\footnote{We can justify (\ref{bc}) as follows. First, we checked (\ref{bc}) diagramatically in perturbation theory, to third order in the background field. The other general argument invokes consideration of the correlation function $
{}_{\rm out}\langle0| \bar\psi_{+} (y^{+},y^{-}) \psi_{+}(x^{+},x^{-})|0\rangle_{\rm in} \, ,
$
where $x^{+}\to -\infty$. In this limit $\psi_{+}(x^{+},x^{-})$  is a free field and we have
$
\psi_+(-\infty, x^{-}) = \sum_{p>0}(a_{p} e^{ipx^{-}}+a_{p}^{\dag}e^{-ipx^{-}})\, .
$
As $a_{p}|0\rangle_{\rm in}=0$ we see that only $e^{-ipx^{-}}$ modes survive. These modes define an
analytic function in $x^-$ in the lower half plane  because $e^{-ipx^{-}}$ decays  when $p>0$ and $\textrm{Im} \,x^{-}<0$. This argument can be applied for any operator $O(\psi_{+})$, to show that a correlation function $\langle ...O(-\infty,x^{-})\rangle$ is analytic in $x^{-}$ in the lower half-plane. }.

Now we return to determining $j_-$ in the expression for the induced current. Working in the retarded gauge $g_{R}(x^{+},x^{-}) \equiv P \exp \int_{-\infty}^{x^{+}} dy^{+} A_{+}(y^{+},x^{-})$ and using (\ref{bc}) one finds
 \begin{align}
j_{-R}(x^-)=-\partial_{-} \Omega_{\rm down} \Omega_{\rm down}^{-1}\,,
\end{align}
where $\Omega_{\rm down}$ and $\Omega_{\rm up}$ are matrices analtyic in the lower and upper half-planes, and solve the matrix Riemann-Hilbert problem
 \begin{align}
\Omega_{\textrm{down}}(x^{-}) \Omega_{\textrm{up}}(x^{-})= P \exp \int_{-\infty}^{+\infty} dy^{+} A_{+}(y^{+},x^{-})\,.
\end{align}
Correspondingly we find $J_{\rm up}(x^-)=-\Omega_{\rm up}^{-1}\partial_{-}\Omega_{\rm up}$ and $J_{\rm down}(x^-)=\partial_{-}\Omega_{\rm down}\Omega_{\rm down}^{-1}$.

Notice that in the spectral gauge (\ref{feyngauge}) we have $j_{-S}(x^{-})=0$. From this it follows that, in the spectral gauge, the effective action is the WZNW action (\ref{easg}), evaluated at $g_S$, and there are no boundary terms. Now, as we determined the current 
\begin{align}
J_{-}=-g_{R}^{-1}\partial_{-}g_{R}-g_{R}^{-1}j_{-R}g_{R}\,,
\end{align} 
one can check that the variation of $W_+(A_{+})$ (see  (\ref{wznwact}) and (\ref{nonAbBound})) indeed equals to (\ref{varofw}).

In the gravitational case everything is similar to the non-Abelian case. In the light-cone coordinates and the axial gauge  $h_{--}=0$, the anomaly equation reads \cite{Polyakov:1987zb}\footnote{To restore the  overall factor in front of  the effective action one needs to replace $ -2 \partial_{-}^{3}h_{++} \to \frac{1}{24\pi}\partial_{-}^{3}h_{++}$.}
\begin{align}
(\partial_{+} -h_{++}\partial_{-}-2(\partial_{-}h_{++})) T_{--} = -2 \partial_{-}^{3}h_{++}\,. \label{gravaneq}
\end{align}
Parametrizing $h_{++}$ by $f(x^{+},x^{-})$, with $(\partial_{+}-h_{++}\partial_{-})f=0$,  the general solution of the equation (\ref{gravaneq}) is 
\begin{align}
T_{--}(x^{+},x^{-})  =-2\mathcal{D}_{-}f+ (\partial_{-}f)^{2} t_{-}(f)\,,
\end{align}
where we define the Schwarzian
\begin{align}
\mathcal{D}_{-}f \equiv \frac{\partial_{-}^{3}f}{\partial_{-}f}- \frac{3}{2}\frac{(\partial_{-}^{2}f)^{2}}{(\partial_{-}f)^{2}}
\end{align}
and $t_{-}(f)$ is at this stage is an arbitrary complex function, which has to be fixed by additional physical arguments\footnote{The logic is very similar to that of the paper \cite{Callan:1992rs}, where the term $t_{-}(f)$ in the stress-energy tensor is fixed by choosing a particular state.}.
So analogously to the non-Abelian case we say that the 
induced current ${}_{\rm out}\langle T_{--}(x^+,x^-) \rangle_{\rm in}$  must satisfy the analytical (spectral) boundary conditions:
\beq\label{bcgrav}
{}_{\rm out}\langle T_{--}(x^+,x^-) \rangle_{\rm in}\to \begin{cases}T_{\rm up}(x^-),& x^+\to+\infty\,,\\ T_{\rm down}(x^-),& x^+ \to -\infty\,,\end{cases}
\eeq
where $T_{\rm up}(x^{-})$ and $T_{\rm down}(x^{-})$ are some complex  functions analytic in the upper and lower $x^-$ half-planes correspondingly. Again, working in the retarded gauge, defined by the condition $f_{R}(x^{+}\to -\infty,x^{-})=x^{-}$ we find that\footnote{It is convenient  here to use the composition  formula for the Schwarzian: $\mathcal{D}_{x}g(f)=\mathcal{D}_{x}f+ (\partial_{x}f)^{2}\mathcal{D}_{f}g\,. $}
\begin{align}
t_{-R}(f)=-2 \mathcal{D}_{f}\Gamma^{-1}_{\rm down}(f)\,,
\end{align}
where $\Gamma_{\rm up}(x^{-})$ and $\Gamma_{\rm down}(x^{-})$ are invertible, analytic functions in the upper and lower $x^-$ half-plane, and they are solutions of the functional Riemann-Hilbert problem ($\Gamma(x^{-})\equiv f_{R}( +\infty,x^{-})$)
\begin{align}
 \Gamma_{\rm down}(\Gamma_{\rm up}(x^{-})) = \Gamma(x^{-})\,.
\end{align}
We have $T_{\rm up}(x^-)=-2\mathcal{D}_{-}\Gamma_{\rm up}$ and  $T_{\rm down}(x^-)=-2\mathcal{D}_{-}\Gamma_{\rm down}^{-1}$ and we again notice that  $t_{-S}(f)=0$ in the spectral gauge $f_{S}$, defined in (\ref{spgg}), which leads to the formula (\ref{effgravspect}). 

Having the expression for the current $T_{--}= -2\mathcal{D}_{-}\Gamma_{\rm down}^{-1}(f_{R})$, we can check that the variation of (\ref{gravefa}) is indeed equal to
\begin{align}
\delta W(h_{++}) = \int d^{2}x\, T_{--}\delta h_{++}\,.
\end{align}

\section{Gauge-gravity duality in two dimensions}\label{app:gauge-gravity} 

In this appendix, we review the duality between 2-dimensional gravity and $ \textrm{SL}(2,\mathbb{C})$ gauge theory~\cite{Alekseev:1988ce,Bershadsky:1989mf,Polyakov:1989dm}.  We find it useful, as the functional Riemann-Hilbert problem can be related to $\textrm{SL}(2,\mathbb{C})$ matrix Riemann-Hilbert problem.\footnote{We need to extend the gauge group to be complex valued, as we are interested in both real and imaginary parts of the action. Originally the duality was found using $\textrm{SL}(2,{\mathbb R})$ gauge group. The only new subtleties arise in treating integrations by parts, but, as long as we use the spectral gauge condition, the formulas are similar to the ones in the literature.}

The main idea is to consider the gauge theory on a nontrivial background, and study one particular component of the gauge field. The gauge field has three flavor indices and two spacetime indices, $A^a_\mu$, $a=+,0,-$ (a new occurrence of $\pm$, unrelated to the others in the paper) and $\mu=+,-$.  Now, instead of fixing the axial gauge $A^a_-=0$, we partially fix the gauge by setting
\beq
A^+_-\equiv T_{--},~~ A^-_-=1,~~A^0_-=0\,.
\eeq
It turns out that the remaining gauge freedom on the component $A^+_-$ acts as the Virasoro generators on a stress tensor $T_{--}$. Thus, there is a beautiful duality between a component of a gauge field and the stress tensor of a certain gravitational theory. To complete the duality, one notices that the anomaly equations for the gauge field $A$ are equivalent to the anomaly equations for a metric $g_{++}$, if we identify the induced current in the gauge theory with the metric in the gravitational theory, $J^-_+=g_{++}$. 

In terms of the action functionals, for the $ \textrm{SL}(2,\mathbb{C})$ non-Abelian gauge theory one can establish a relation 
\begin{align}
W_{\rm WZ}(h)= W_{\rm gWZ}(g_{++})\,,
\end{align}
where the Wess-Zumino action and gravitational Wess-Zumino actions are given by the formulas
\begin{align} 
W_{\rm gWZ}(g_{++}) &= \frac{1}{4}\int d^{2}x \left({\partial_{-}^{2}f \partial_{+}\partial_{-}f \over (\partial_{-}f)^{2}}-{(\partial_{-}^{2}f)^{2}\partial_{+}f \over (\partial_{-}f)^{3}}\right)\,, \notag\\
W_{\rm WZ}(h)&= \frac{1}{2} \int_{0}^{1}dt d^{2}x\, \Tr (h^{-1}\dot{h} [h^{-1}\partial_{-}h, h^{-1}\partial_{+}h])\,,
\end{align}
and the $ \textrm{SL}(2,\mathbb{C})$ matrix $h(x^{+},x^{-},t)$ and the metric $g_{++}(x^{+},x^{-})$ are related as follows:
\begin{align}
&A_-=h^{-1}\partial_{-}h =   \left(\begin{array}{cc}
    0 & T_{--} \\ 
    1 & 0 \\ 
  \end{array}\right),\quad J^-_+ =(-h^{-1}\partial_{+}h)_{21}=g_{++},\quad (\partial_{+}-g_{++}\partial_{-})f=0\,,\notag\\
&\qquad\partial_{+}T_{--}-g_{++}\partial_{-}T_{--}-2(\partial_{-}g_{++})T_{--}=-\frac{1}{2}\partial_{-}^{3}g_{++}\,.
\end{align}
One can prove these relations using a nice parametrization for the matrix $h$:
\begin{align}
h= \left(\begin{array}{cc}
    a & \partial_{-}a \\ 
    b & \partial_{-}b \\ 
  \end{array}\right),\quad \textrm{with} \quad a\partial_{-}b-b\partial_{-}a=1\,.
\end{align}
In this parametrization one has   $g_{++}= a\partial_{+}b-b\partial_{+}a$ and $T_{--}=\partial_{-}^{2}a/a=\partial_{-}^{2}b/b$ and $f= \mathcal{F}(a/b)$, where $\mathcal{F}$ is an arbitrary invertible function. Thus, in terms of $h$, we can find the characteristic function $f$. It is inetersting to understand whether this makes  a connection between functional and matrix Riemann-Hilbert problems.

\section{Non-Abelian and gravitational corrections to Caldeira-Leggett formula}\label{CaldLeg}
In the case of a weak non-Abelian field profile, we may try to solve the matrix Riemann-Hilbert problem perturbatively 
\begin{align}
\Omega_{\rm down}(x^{-})\Omega_{\rm up}(x^{-}) = \Omega(x^{-})\,,
\end{align}
where $\Omega(x^{-}) \equiv P \exp \int_{-\infty}^{\infty} A_{+}(y^{+},x^{-}) dy^{+}$ and we assume the following perturbative decomposition for  $\Omega_{\rm down}$
and $\Omega_{\rm up}$:
\begin{align}
&\Omega_{\rm up}=\mathbbm{1}+\Omega_{\rm up}^{(1)}+\Omega_{\rm up}^{(2)}+\dots, \quad \Omega_{\rm down}=\mathbbm{1}-\Omega_{\rm down}^{(1)}-\Omega_{\rm down}^{(2)}+\dots\,.
\end{align}
Expanding $\Omega(x^{-})$ to first order  we get 
\begin{align}
\Omega^{(1)}_{\rm up}(x^{-})-\Omega^{(1)}_{\rm down}(x^{-})= \omega(x^{-})\,,
\end{align}
where $\omega(x^{-})\equiv\int_{-\infty}^{\infty} A_{+}(y^{+},x^{-}) dy^{+}$,
thus 
\begin{align}
\Omega_{\rm up}^{(1)} = \omega_{\rm up}(x^{-}), \quad \Omega_{\rm down}^{(1)}  =\omega_{\rm down}(x^{-})\,, \label{g1pert}
\end{align}
where $ \omega_{\rm up/down}(x^{-})$ are given in (\ref{omupdown}). 
At second order we have 
\begin{align}
\Omega_{\rm up}^{(2)}-\Omega_{\rm down}^{(2)}  =\int_{-\infty}^{+\infty}dy_{1}^{+}\int_{-\infty}^{y_{1}^{+}}dy_{2}^{+}\, \Tr (A_{+}(y_{1}^{+},x^{-})A_{+}(y_{2}^{+},x^{-}))+\omega_{\rm down}\omega_{\rm up}\,,
\end{align}
where we used (\ref{g1pert}), and so we have just a scalar Riemann-Hilbert problem, which we and can solve explicitly. 
Now plugging this perturbative decomposition in the $2$-form (\ref{nonAbBound}) we obtain
\begin{align}
\textrm{Im}\,W_{B}(\Omega_{\rm up},\Omega_{\rm down}) &= \,\textrm{Im} \int dx^{-} \, \Tr \Big(\omega_{\rm down} \partial_{-}\omega_{\rm up} +\notag\\&+ \Omega_{\rm down}^{(2)}\partial_{-}\omega_{\rm up}+\omega_{\rm down}\partial_{-}\Omega_{\rm up}^{(2)}-\frac{1}{2}\big(\omega_{\rm up}\partial_{-}\omega_{\rm down}^{2}+\omega_{\rm down} \partial_{-}\omega_{\rm up}^{2}\big)+\dots\Big)\,,
\end{align}
where the term in the first line is the standard Caldeira-Leggett formula, and the terms in the second line are the first perturbative corrections to it, cubic in $A_+$. Notice that,  perturbatively, it is clear that the imaginary part of $W_B$ does not depend on the $t$-interpolation.

Now, in analogy with  the non-Abelian case, we can solve the functional Riemann-Hilbert problem perturbatively. This assumes that the gravitational field is weak. It is convenient to write $\Gamma_{\rm up/down}(x^{-})=x^{-}\pm \gamma_{\rm up/down}(x^{-})$ and $\Gamma(x^{-})=x^{-}+\gamma(x^{-})$, then for  (\ref{frhp}) we have
\begin{align}
\gamma_{\rm up}(x^{-})- \gamma_{\rm down}(x^{-}+\gamma_{\rm up}(x^{-}))=\gamma(x^{-})\,.
\end{align}
Then writing a perturbative decomposition for $\gamma_{\rm up/down}$
\begin{align}
\gamma_{\rm up}= \gamma_{\rm up}^{(1)}+ \gamma_{\rm up}^{(2)}+\dots, \quad 
\gamma_{\rm down}= \gamma_{\rm down}^{(1)}+ \gamma_{\rm down}^{(2)}+\dots
\end{align}
we find at the first and the second order 
\begin{align}
&\gamma_{\rm up}^{(1)}(x^{-})-\gamma_{\rm down}^{(1)}(x^{-})= \gamma(x^{-})\,,\notag\\
&\gamma_{\rm up}^{(2)}(x^{-})-\gamma_{\rm down}^{(2)}(x^{-}) =\gamma_{\rm up}^{(1)}(x^{-})\partial_{-}\gamma_{\rm down}^{(1)}(x^{-})\,.
\end{align}
So we see that step by step we just need to solve the scalar Riemann-Hilbert problem, which has the explicit solution (\ref{omupdown}). Thus, the boundary action (\ref{gbacccc}) reads
\begin{align}
W_{B} = \int dx^{-}\Big(\gamma_{\rm down}^{(1)}\partial_{-}^{3}\gamma_{\rm up}^{(1)}-\big((\partial_{-}\gamma_{\rm up}^{(1)})^{2}\partial_{-}^{2}\gamma_{\rm down}^{(1)}+(\partial_{-}\gamma_{\rm down}^{(1)})^{2}\partial_{-}^{2}\gamma_{\rm up}^{(1)}-(\partial_{-}^{2}\gamma_{\rm down}^{(1)})^{2}\gamma_{\rm up}^{(1)}\big)+\dots\Big)\,.
\end{align}
We also checked this result using Feynman diagrams.

\addcontentsline{toc}{section}{References}
\bibliographystyle{utphys}
\bibliography{Refs}

\end{document}